\documentclass[utf8]{FrontiersinHarvard} % for articles in journals using the Harvard Referencing Style (Author-Date), for Frontiers Reference Styles by Journal: https://zendesk.frontiersin.org/hc/en-us/articles/360017860337-Frontiers-Reference-Styles-by-Journal
\usepackage{url,hyperref,lineno,microtype,subcaption}
\usepackage[onehalfspacing]{setspace}
\def\keyFont{\fontsize{8}{11}\helveticabold }
\def\firstAuthorLast{} 
\def\Authors{Daryoush Shiri\,$^{1,*}$, Reza Nekovei\,$^{2}$ and Amit Verma\,$^{2}$}
\def\Address{$^{1}$Department of Microtechnology and Nanoscience, Chalmers University of Technology, Gothenburg, Sweden \\
$^{2}$Department of Electrical and Computer Engineering, Texas A\&M University-Kingsville, Kingsville, Texas, USA}
\def\corrAuthor{Daryoush Shiri}
\def\corrEmail{shiri@chalmers.se}
\begin{document}
\onecolumn
\firstpage{1}
\title[]{Low-Temperature Electron Transport in [110] and [100] Silicon Nanowires: A DFT - Monte Carlo study} 

\author[\firstAuthorLast ]{\Authors} %This field will be automatically populated
\address{} %This field will be automatically populated
\correspondance{} %This field will be automatically populated
\extraAuth{}% If there are more than 1 corresponding author, comment this line and uncomment the next one.
%\extraAuth{corresponding Author2 \\ Laboratory X2, Institute X2, Department X2, Organization X2, Street X2, City X2 , State XX2 (only USA, Canada and Australia), Zip Code2, X2 Country X2, email2@uni2.edu}

\maketitle

\begin{abstract}

%%% Leave the Abstract empty if your article does not require one, please see the Summary Table for full details.
\section{}
The effects of very low temperature on the electron transport in a [110] and [100] axially aligned unstrained silicon nanowires (SiNWs) are investigated. A combination of semi-empirical 10-orbital tight-binding method, density functional theory (DFT), and Ensemble Monte Carlo (EMC) methods are used. Both acoustic and optical phonons are included in the electron-phonon scattering rate calculations covering both intra-subband and inter-subband events. A comparison with room temperature (300 K) characteristics shows that for both nanowires, the average electron steady-state drift velocity increases at least 2 times at relatively moderate electric fields and lower temperatures. Furthermore, the average drift velocity in [110] nanowires is 50 percent more than that of [100] nanowires, explained by the difference in their conduction subband effective mass. Transient average electron velocity suggests that there is a pronounced streaming electron motion at low temperature which is attributed to the reduced electron-phonon scattering rates. 

\tiny
 \keyFont{\section{Keywords:} Silicon Nanowire, cryogenic, electron-phonon scattering, DFT, Ensemble Monte Carlo, CMOS, spin qubit} %All article types: you may provide up to 8 keywords; at least 5 are mandatory.
\end{abstract}

\section{Introduction}

Since the first implementation of top-down \cite{Singh2008} and bottom-up \cite{Ma2003} approaches to fabricate Silicon nanowires (SiNWs), they have constantly shown promising applications in different areas of technology. These are all fueled by the compatibility of their fabrication with the mainstream silicon technology and enhanced quantum mechanical effects as a result of size reduction \textit{\textit{e.g.,}} direct bandgap. Tunability of the optical absorption and direct nature of the the bandgap opened SiNWs way into the photonic realm for example photo-detectors, resolving photon polarization \cite{Park2015photodet,Zhao2017}, and photovoltaic \cite{Gonchar2019}. The surface effects in SiNWs lead to more sensitivity for chemical sensors \cite{Kashyap2022}. Breaking the centro-symmetry of the crystal in SiNWs due to strain or surface effects enhances the nonlinear optical effects \textit{e.g.,} second harmonic generation \cite{Weicha2015} and third order nonlinear effects \cite{Park2023nonlin}.

SiNWs have also shown promising benefits in enhancing the coherence of spin-based quantum bits (qubits) as opposed to III-V nanowires in which the coherence of the qubits is limited due to hyperfine magnetic interaction with nuclei. Implementing spin-based qubit chips based on CMOS-compatible SiNW systems are on the rise \cite{Zwanenburg2009, Maurand2016, Hu2012, Piot2022}.

Low temperature effects on the charge carrier transport in silicon nanowires also opens up new horizons in understandings and possible low-temperature (cryogenic) applications \textit{e.g.,} CMOS-compatible cryogenic sensors, switches, and deep-space electronics \cite{Jones2020,Rohrbacher2023}. 
For the latter, the traditionally large bandgap III-V semiconductors are of use despite the high price of their wafer fabrication and processing. SiNWs with direct and controllable bandgap promise a low-cost alternative for III-V counterparts. Different  electronic applications of SiNWs are also rising \cite{Arjmand2022,Schmidt2009}, thanks to the developments of CMOS-compatible top-down fabrication methods \cite{Singh2008,Pott2008}. 

In this article we have studied the effect of low temperature on the transport of electrons at both steady-state and transient conditions under the influence of electron-phonon scattering. The scattering events include both intra- and inter-subband transition processes due to longitudinal acoustic (LA) and longitudinal optical (LO) phonons. Two SiNWs of different crystallographic directions were chosen for the study: a 1.3nm [110] and a 1.1nm [100] SiNWs terminated with hydrogen atoms. The band structure data (conduction subbands for electrons) and scattering rates are used by an ensemble Monte Carlo (EMC) code for calculation of electron transport under the influence of applied electric field. The EMC method is a very useful tool to investigate steady-state and transient phenomena in semiconducting nano-devices \cite{Tomizawa1993}. 

In the next section, we discuss the computational methods including the calculation of band structures, electron-phonon scattering rates and EMC methods. In section 3 we discuss the results. We show that at low temperatures the average drift velocity is enhanced by at least a factor of two due to mitigation of scattering events involving phonon absorption. The difference in the effective mass of [110] and [100] SiNWs leads to better transport (higher electron velocity) in [110] SiNWs. Finally, we show that the initial back-and-forth displacement of electron population in the momentum space, within the first Brillouin zone (BZ), causes ringing or streaming motion at moderate bias electric fields. This is supported through transient EMC snapshots of the electron population at different time scales. This effect as well as saturation of drift velocity at higher fields corroborates with previous studies in carbon nanotubes (CNTs) \cite{Jovanovic1992,Ahmadi2008}.  

\section{Computational Methods}

\subsection{Energy minimization and Band structure} Two silicon nanowires of different crystallographic directions are investigated here: [110] and [100]. The average diameters for these nanowires are 1.3 nm, 1.1 nm, respectively. The nanowires are considered freestanding, with the surface silicon dangling bonds passivated with hydrogen atoms. Terminating the dangling Si atoms on SiNW is a model for a nanowire surrounded by ideal vacuum or a large-bandgap material or a perfect oxide cladding free of dangling bonds or dislocations. The structural energy of the nanowires is minimized using the density functional theory code in SIESTA. This process allows for the most energetically favorable unit cell to take form and the formation of dangling $\rm SiH$ and $\rm SiH_2$ groups on the SiNW edge are finalized \cite{Soler2002}. 

\begin{figure}[h!]
\begin{center}
\includegraphics[width=14cm]{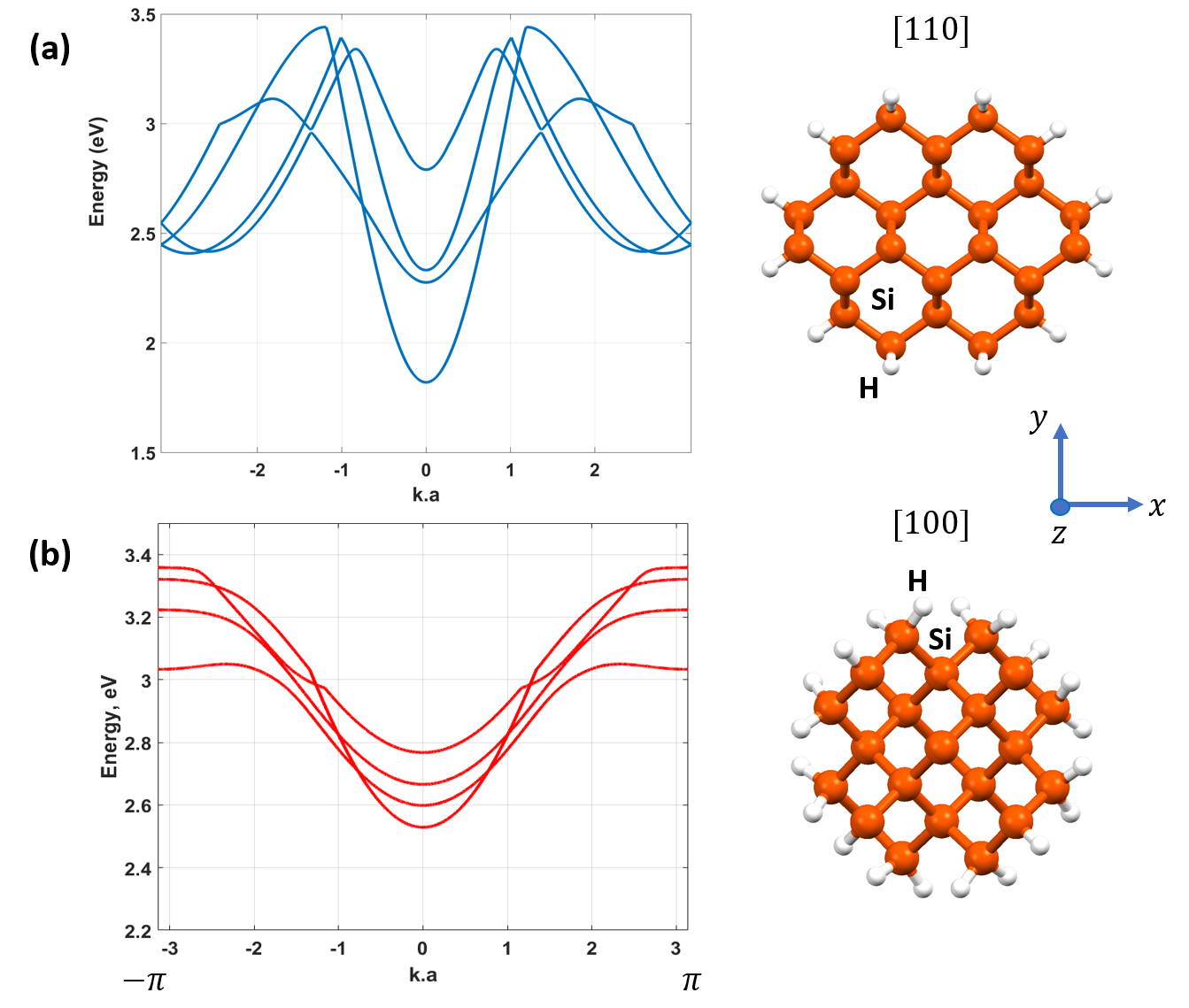}% This is a *.eps file
\end{center}
\caption{(a) The band structure of 1.3 nm [110] SiNW showing the first four conduction subbands used in the electron-phonon scattering calculation. (b) The band structure showing the first four conduction subbands for 1.1 nm [100] SiNW. The $xy$ cross sections of the SiNWs are shown on the right side. The orange and white atoms are Si and H, respectively.}
\label{fig:1}
\end{figure}

The exchange-correlation functional which was used is the Generalized Gradient Approximation (GGA) type with Perdew-Burke-Ernzerhof (PBE) pseudopotentials. The number of $k$-points to sample the Brillouin Zone (BZ) are 1×1×40 based on Monkhorst-Pack algorithm with many number of points along the periodical axis of the nanowire (z axis). The minimum distance of adjacent unit cells is more than 0.6 nm to avoid any possible wavefunction overlapping. The energy cut-off, split norm, and force tolerance are 680 eV, 0.15, and 0.01 eV/A°, respectively. The energy of the unit cell of nanowires is minimized using the conjugate gradient (CG) algorithm with a variable unit cell option. This option allows the volume of the unit cell to grow or shrink depending on the movement of Si-Si and Si-H bonds, particularly the canting of the silicon dihydride groups ($\rm SiH_2$) on the surface of SiNW. Figure \ref{fig:1} (a) and (b) show the band structure and $xy$ cross sections of hydrogen-passivated SiNWs in [110], and [100] directions, respectively. \\
After obtaining the coordinates of the atoms for the energy-minimized unit cells, the band structure of nanowires are calculated with a semi-empirical $sp^3d^5s^*$ tight-binding (TB) scheme using parameters given by \cite{Jancu1998}. It was shown that the TB method can faithfully reproduce the experimental data of photoluminescence in silicon nanowires under strain as reported in \cite{Walavalkar2010,Demichel2011,Bae2018}. As can be seen in Figure \ref{fig:1}.(a), both nanowires are of direct bandgap type. In the [110] SiNW the minimum conduction subband energy is $E_{cmin} =1.81~\text{eV}$. In the [100] SiNW the minimum of the conduction band is at $E_{cmin} = 2.528~\text{eV}$. The effect of quantum confinement is more pronounced for the [100] SiNW, \textit{i.e.,} as it is narrower it has a higher energy level. The effective mass of [100] subbands is four times that of [110] (\textit{e.g.,} $m^*=0.16$ for [110] and $m^*=0.63$ for [100] for the first subband). As we will see later this leads to a less average drift velocity for [100] SiNWs.

\subsection{Phonon scattering rates and EMC}
For the calculation of electron-phonon scattering rates, four first subbands are chosen. The reason for this is to ensure that the energy difference between the minimum conduction energy and the highest one is more than $5k_BT$ which is $130~\text{meV}$ at room temperature \textit{i.e.,} $T = 300~\text{K}$. The 4th subband minimum is at $E_{c4}= 2.8~\text{eV}$ and $E_{c4}= 2.76~\text{eV}$ for [110] and [100] SiNWs, respectively. These four subbands are included in the EMC calculations.
For the electron-phonon scattering rate calculation, the 1st Brillouin Zone (BZ) is divided into 8000 grid points along $k$ direction. For each $k$ point, the subband energies and eigenstates (TB wave functions) are calculated and tabulated. The electron-phonon scattering rates are numerically evaluated using the 1st order perturbation theory and deformation potential approximation, for different temperatures. Both types of phonons \textit{i.e.,} longitudinal acoustic (LA) and optical (LO) phonons are included. The electron scattering includes both intra-subband and inter-subband scattering events. For acoustic phonons, we used the Debye approximation, \textit{i.e.,} it is assumed that the acoustic phonons have linear dispersion and their energy ($E_P=\hbar\omega$) and wave vector ($|k|$) are linearly proportional. Therefore, $E_P=\hbar\omega=c|k|$, where $\hbar$ is Planck’s constant and $c$ is the velocity of sound in silicon, and $\omega$ is the acoustic phonon frequency. For LO phonons, the dispersion is almost flat and as a result, all optical phonons can be considered to have the same energy which for silicon is $E_{P}=E_{LO}= 54~\text{meV}$. The scattering rates and the indices of the possible final (secondary) states in the BZ (after scattering) are sorted in a table depending on if they are phonon-absorption or phonon-emission type and if they are intra- or inter-subband. Indices of the secondary states to which electron scatters by absorption or emission of a phonon is used by the EMC algorithm to decide how an electron injected into the nanowire propagates as a result of applied electric field. Details of the scattering rates calculations are presented in \cite{Buin2008,Shiri2018}. 

The total scattering rates from the 1st subband to other bands (the 1st one included), due to LO and LA phonons for different nanowires and different temperatures are compared in Figure \ref{fig:2}. Figure \ref{fig:2a} shows the total scattering rates due to LO and LA phonons at room temperature calculated for [100] and [110] nanowires. As can be seen the scattering rates in [100] SiNW are overall higher by a factor of two compared to the [110] SiNW which can be attributed to the effective mass difference of these two nanowire directions as it was shown in Figure \ref{fig:1}.\\ This is because the scattering rate depends on the availability of the secondary states quantified by the density of states (DOS). In a 1D solid like nanowire, DOS is proportional to the square root of the effective mass \textit{i.e.,} $DOS(E) \propto \sqrt{m^*}$.
The scattering rates from the 1st subband to other subbands at two extreme temperatures of $T = 4~\text{K}$ and $T = 300~\text{K}$ are compared in Figure \ref{fig:2b} for [100] nanowire. This is similar to the observations for [110] silicon nanowires as reported in \cite{AmitNMDC23}. The higher scattering rate at higher temperatures is due to higher phonon absorption scattering. The peaks in the LO scattering rates emanate from the van Hove singularities. They correspond to electron transitions to the bottom of different subbands once the electron energy reaches the $E_{LO}= 54~\text{meV}$ which is onset for a LO phonon emission event to occur. On the contrary, the peaks of LA scattering events are smooth as in the calculation of scattering rates a continuum of secondary states is available for every energy of electron due to linear dispersion of the acoustic phonons. 
At low temperatures, the dominant scattering event is due to the emission of LO phonons as well as acoustic phonons. The observed LO phonon peaks have significant effect on the electron transport under both steady-state and transient conditions, as will be seen shortly.
The SiNWs are defect-free, infinitely long, and undoped. The temperature is assumed to be uniform. The applied electric field is along the nanowire axis ($z$ axis) and is uniform. For the steady-state analysis, the electrons are injected at $t = 0~\text{s}$ at the bottom of the lowest conduction band. For the transient results, the simulation is first run for 50,000 iterations at an electric field of $E_{field} = 0~\text{kV/cm}$ to achieve a near equilibrium distribution. We use the standard EMC algorithm and methodology as presented by \cite{Jacoboni1989} in our simulations. 

\section{Results and discussions}
Figure \ref{fig:3} shows the average electron drift velocity as a function of the applied electric field along the length of the nanowires ($z$ axis) for $T = 4~\text{K}$ and $T = 300~\text{K}$. As expected, the drift velocity is significantly higher for the lower temperatures due to reduced scattering rates which were shown in Figure \ref{fig:2}. At an electric field of $E_{field}=15~\text{kV/cm}$, the drift velocity at [100] drops approximately by one fifth and for [110] it drops by a half as the temperature is increased to 300~K. This is attributed to the higher total scattering rates at higher temperatures. The enhancement of transport at $T=4~\text{K}$ is also observed in the IV characteristics of cryogenic gate-all-around SiNW FETs reported by \cite{Rohrbacher2023}. 
The electron-phonon scattering rate is proportional to $n(E_P)+1$ and $n(E_P)$ for phonon emission and absorption events, respectively. The population of phonons, $n(E_P)$, is determined by the Bose-Einstein factor and is given as:
\begin{equation}
    n(E_P)=\frac{1}{e^{E_P/{k_BT}}-1}
\end{equation}
where the phonon energy is $E_P$ and $k_B = 1.3807\times 10^{-23} ~\text{J/K}$ is Boltzmann constant. As the temperature is lowered, the predominant scattering event becomes phonon emission because $n(E_P)\rightarrow 0$. As can be seen in Figure \ref{fig:3}, the average drift velocity saturates at high electric fields. This is because velocity saturation occurs primarily through phonon emission scattering and is nearly temperature-independent threshold process \cite{Verma2009}. It is also important to look at the transient distribution function in conjunction with the scattering rates in Figure \ref{fig:2}. Figure \ref{fig:4} depicts the evolving (time-dependent) electron distribution function at $T = 300~\text{K}$ for the applied electric field of $30~\text{kV/cm}$. As can be seen, the distribution function appears to stop moving beyond a wave vector $k$ value that roughly corresponds to a peak in the scattering rates at approximately $k = 2\times10^6~\text{1/cm}$ for [110] and $k = 6\times10^6~\text{1/cm}$ for [100] (see red plots in Figure \ref{fig:4}). These $k$ values correspond to blue and red peaks in LO data of Figure \ref{fig:2} and prove that the LO-phonon emission scattering peak impedes electrons from gaining relatively large crystal momentum with an increase in the electric field. \\
\indent The initial back and forth displacement of electron population in the BZ can be appreciated if we plot the time evolution of average drift velocity which reveals a wealth of information about the scattering mechanism and shows the importance of the peaks in scattering rates through phonon emission. Figure \ref{fig:5} shows the transient average electron drift velocity for the temperatures considered at an electric field of $E_{field}= 20~\text{kV/cm}$. Firstly, it shows that the velocity is enhanced by decreasing the temperature which is again due to lower scattering rates through phonon absorption at lower temperatures. \\
%Secondly, it shows that the velocity difference between [110] and [100] SiNWs is more pronounced at low temperatures (a factor of 4 at $T = 4~\text{K}$) as opposed to higher temperatures (a factor of 2 at $T = 300~\text{K}$).
\indent Secondly, the streaming motion of electrons \textit{i.e.,} the ringing of the velocity at initial times becomes more pronounced at lower temperatures, which is also predicted by \cite{Jovanovic1992}. The streaming motion can be understood through the transient evolution of the electron distribution function (Figure \ref{fig:4}) and scattering rates as shown in Figure \ref{fig:2}. The reason for the velocity oscillation in Figure \ref{fig:5} (\textit{e.g.,} from 0 to 600 fs) can be understood if we look at electron distribution inside the 1st BZ versus time as it was shown in Figure \ref{fig:4}. The distribution function is seen to ‘bounce’ back and forth. In short, once the electric field is applied, electrons quickly gain crystal momentum and reach the first phonon emission scattering peak. This is where electrons are $E_{LO} = 54~\rm meV$ above the minimum of the first conduction subband (approximately $k = 2\times10^6~\text{1/cm}$ for [110] and $k = 6\times10^6~\text{1/cm}$ for [100]). At this point, a significant number of electrons lose momentum and fall back to near the $k = 0$ or BZ center, where they accelerate again. The process continues until phonon absorption scattering events cause the electrons to reach an average steady-state drift velocity. Saturation of the velocity at higher electric fields also corroborates with the electron transport under electric fields in narrow CNTs in \cite{Ahmadi2008}. Recall that the scattering rates for emission and absorption of phonon are proportional to $n(E_P)+1$ and $n(E_P)$, respectively. Hence, at very low temperatures the emission term is dominant as the absorption term approaches to zero \textit{i.e.,} $n(E_P)\rightarrow 0$. At high temperatures, however, the difference between the phonon populations becomes very small \textit{i.e.,} $n(E_P)+1\approx n(E_p)$, which means both emission and absorption rates are dominant and the average of drift velocity settles to a lower but stable value relatively faster.

\section{Conclusions}
Using DFT and Ensemble Monte Carlo methods we demonstrated that low temperature has significant effects on electron drift velocity in [110] and [100] silicon nanowires. It was shown that the velocity is enhanced at $T = 4~\text{K}$ in both nanowires as a result of diminished electron-phonon absorption scattering. For both nanowires, the velocity saturates at high electric fields showcasing the dominance of scattering events due to LO-phonons. This was examined by looking into the time-dependent bounce of electron population within the 1st BZ which also explained the observed streaming motion of electrons particularly at lower temperatures.  Also we demonstrated that [110] SiNWs have better transport properties than [100] SiNWs. This is due to higher effective mass in [100] SiNWs which manifest itself in higher DOS for electron secondary states and, as a result, increased electron-phonon scattering rate. These observations promise potential applications of SiNWs at low temperatures \textit{e.g.,} cryogenic devices and circuits. 

\section*{Conflict of Interest Statement}
The authors declare that the research was conducted in the absence of any commercial or financial relationships that could be construed as a potential conflict of interest.

\section*{Author Contributions}

Daryoush Shiri did DFT and TB simulations and calculated the electron-phonon scattering data. Amit Verma developed the basic EMC codes to perform the charge transport calculations. Reza Nekovei adapted and deployed EMC codes for different compilers and environments on the Computing Clusters. All authors contributed to the data analysis and writing the manuscript. 

%\section*{Funding}
%Details of all funding sources should be provided, including grant numbers if applicable. Please ensure to add all necessary funding information, as after publication this is no longer possible.

\section*{Acknowledgments}
The authors sincerely acknowledge the access to the Stampede2 Supercomputing machine provided by the Texas Advanced Computing Center (TACC) at Austin, U.S.A.

%\section*{Data Availability Statement}
%The datasets [GENERATED/ANALYZED] for this study can be found in the [NAME OF REPOSITORY] [LINK].
% Please see the availability of data guidelines for more information, at https://www.frontiersin.org/about/author-guidelines#AvailabilityofData

\bibliographystyle{Frontiers-Harvard} %  Many Frontiers journals use the Harvard referencing system (Author-date), to find the style and resources for the journal you are submitting to: https://zendesk.frontiersin.org/hc/en-us/articles/360017860337-Frontiers-Reference-Styles-by-Journal. For Humanities and Social Sciences articles please include page numbers in the in-text citations 
\bibliography{LowTemp_transport_in_100110SiNWs_Sept11_2024}

%%% Make sure to upload the bib file along with the tex file and PDF
%%% Please see the test.bib file for some examples of references

\section*{Figures}

%------------------------------- FIGURES --------------------------------------------
%-----------------------------------------------------------------------------------------
\setcounter{figure}{2}
\setcounter{subfigure}{0}
\begin{subfigure}
\setcounter{figure}{2}
\setcounter{subfigure}{0}
    \centering
    \begin{minipage}[b]{0.5\textwidth}
        \includegraphics[width=\linewidth]{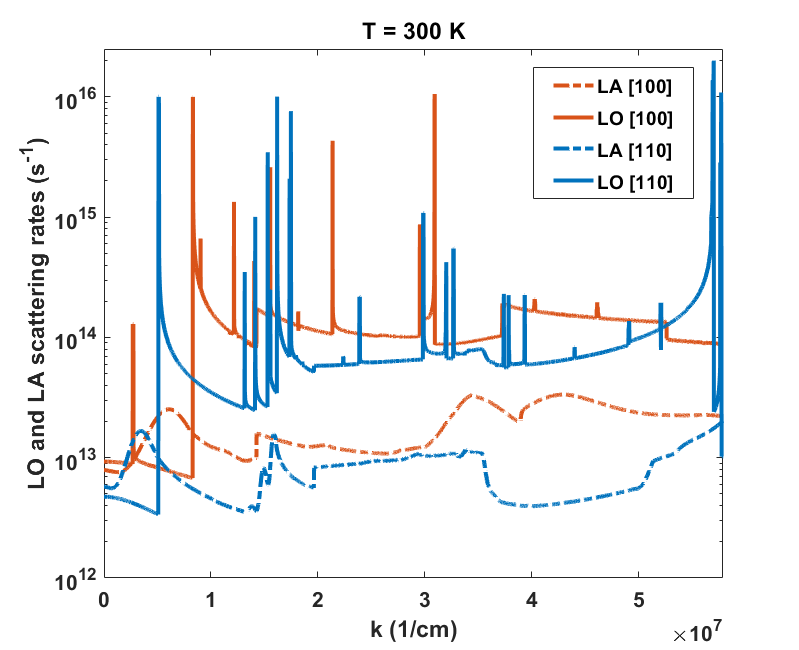}
        \caption{}
        \label{fig:2a}
    \end{minipage}  
   
\setcounter{figure}{2}
\setcounter{subfigure}{1}
    \begin{minipage}[b]{0.5\textwidth}
        \includegraphics[width=\linewidth]{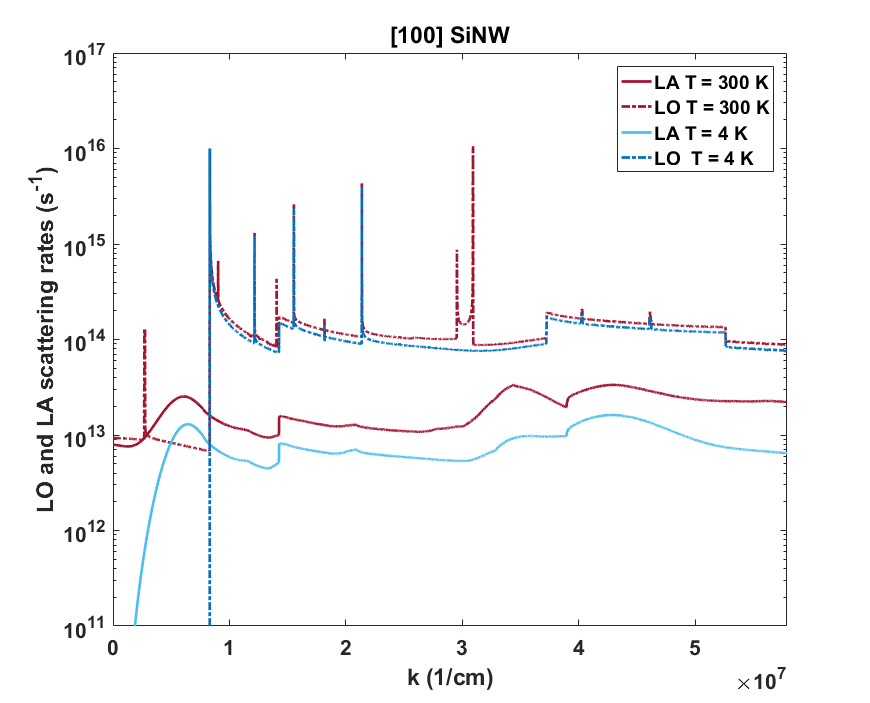}
        \caption{}
        \label{fig:2b}
    \end{minipage}

\setcounter{figure}{2}
\setcounter{subfigure}{-1}
    \caption{(a) The electron-phonon scattering rates for LA and LO phonons at $T = 300~\text{K}$. The [110] nanowire (blue plots) has lower scattering rates compared to [100] (red plots).(b) The comparison of LO and LA scattering rates for [100] SiNW at $T = 4~\text{K}$ and $T =300~\text{K}$ temperatures. The scattering rates are enhanced by increasing the temperature as shown by crimson plots.}
    \label{fig:2}
\end{subfigure}
%----------------------------------------------------------------------------------------------

\begin{figure}[h!]
\begin{center}
\includegraphics[width=10cm]{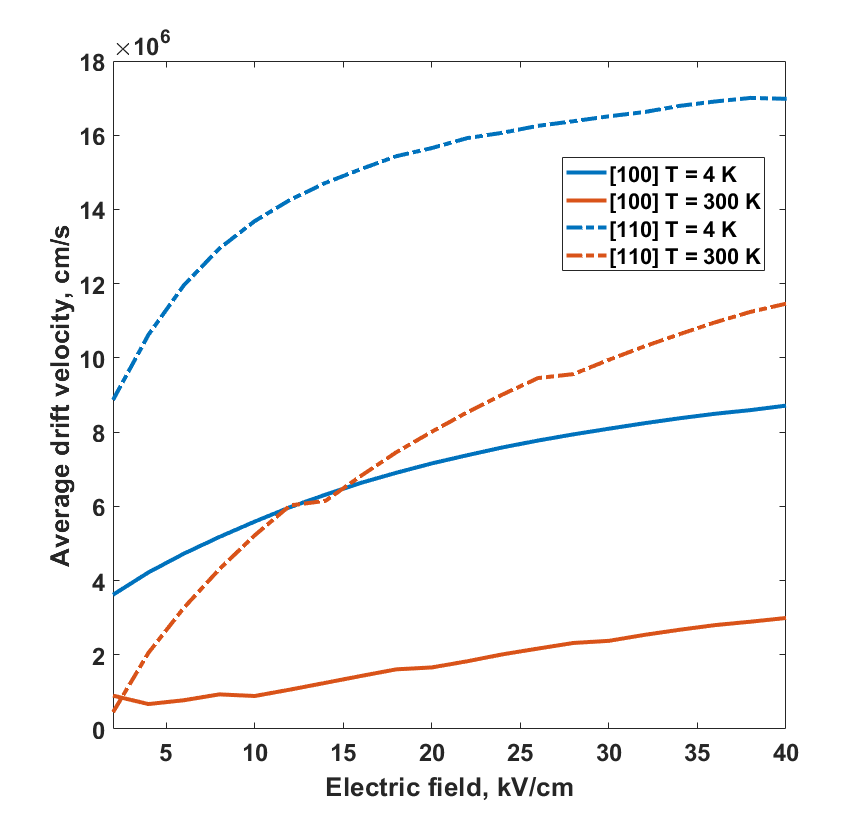}% This is a *.eps file
\end{center}
\caption{ The average drift velocity of electron as a function of applied longitudinal electric field for [110] (dashed line) and [100] (solid line) SiNWs at $T = 4~\text{K}$ (blue) and $T = 300~\text{K}$ (red). Higher velocity (less scattering) is visible for [110] SiNW (dashed plots).}
\label{fig:3}
\end{figure}

\setcounter{figure}{4}
\setcounter{subfigure}{0}
\begin{subfigure}
\setcounter{figure}{4}
\setcounter{subfigure}{0}
    \centering
    \begin{minipage}[b]{0.5\textwidth}
        \includegraphics[width=\linewidth]{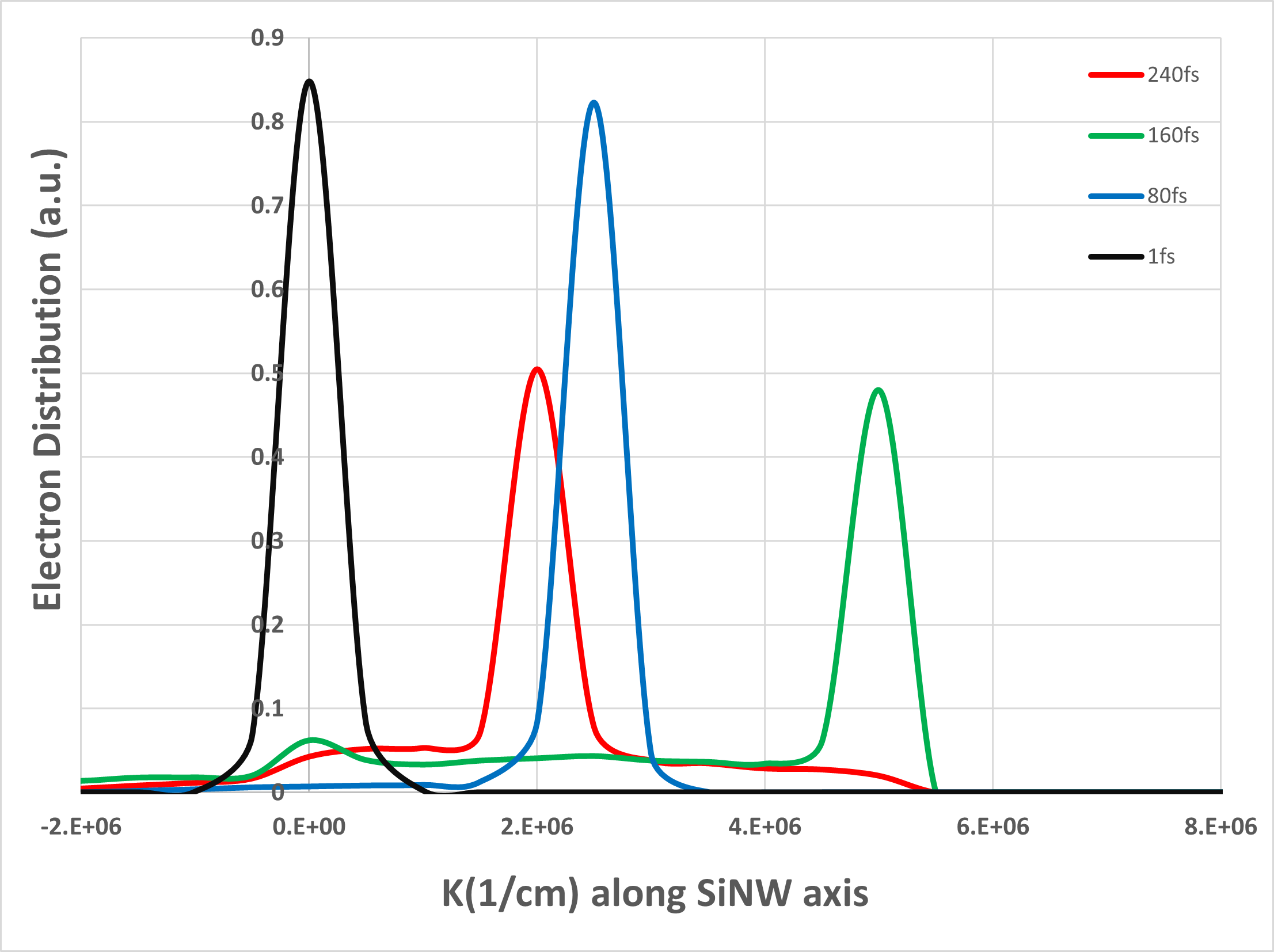}
        \caption{}
        \label{fig:4a}
    \end{minipage}  
   
\setcounter{figure}{4}
\setcounter{subfigure}{1}
    \begin{minipage}[b]{0.5\textwidth}
        \includegraphics[width=\linewidth]{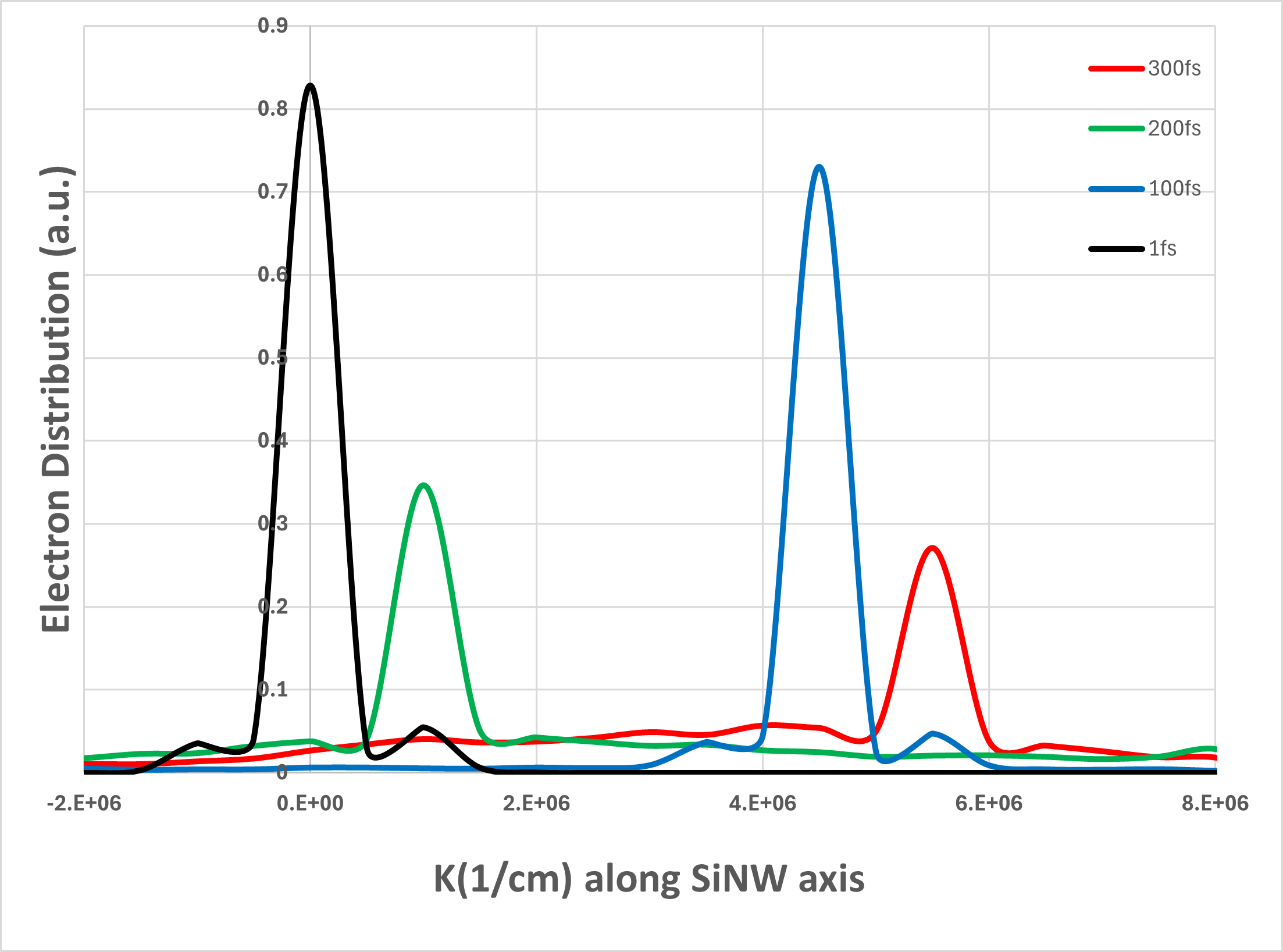}
        \caption{}
        \label{fig:4b}
    \end{minipage}

\setcounter{figure}{4}
\setcounter{subfigure}{-1}
    \caption{(a) Oscillation of electron distribution within the 1st BZ at a constant electric field ($E_{field}= 30~\text{kV/cm}$) at $T = 4~\text{K}$ for [110] SiNW. (b) The same data as (a) for [100] SiNW at $T= 4~\text{K}$. The red plots show that as time passes the distribution is mostly centered around the $k$-points which correspond to LO peaks in Figure~\ref{fig:2}.}
    \label{fig:4}
\end{subfigure}

\begin{figure}[ht]
\begin{center}
\includegraphics[width=10cm]{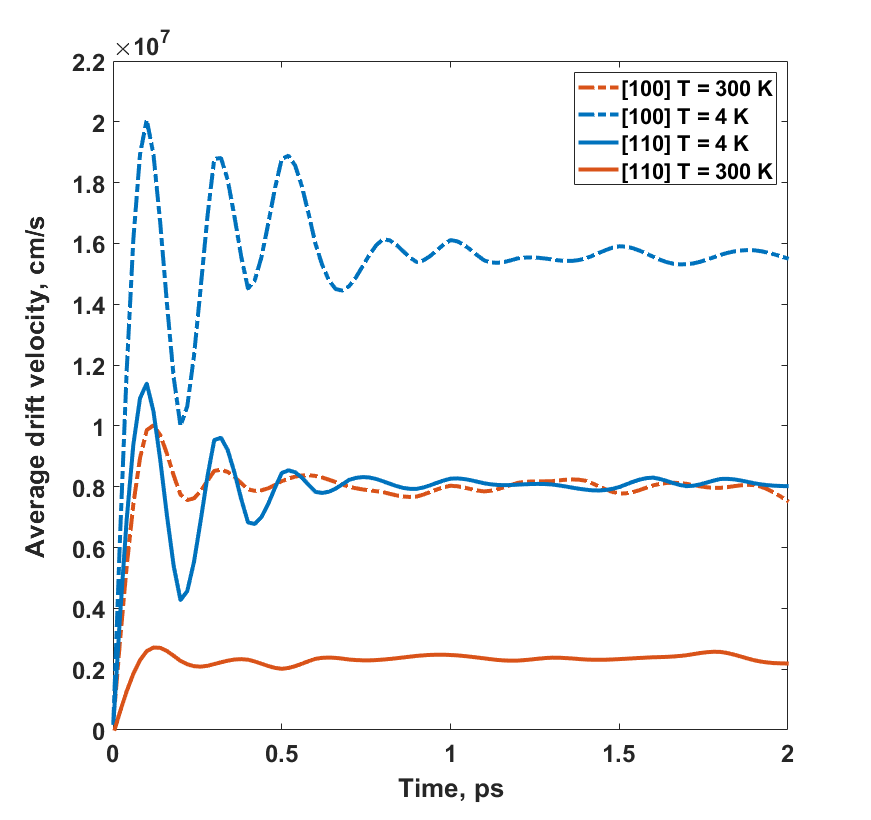}% This is a *.eps file
\end{center}
\caption{Time evolution of electron average drift velocity for [110] and [100] SiNWs at low and high temperatures. The oscillation of velocity is more pronounced at the low temperature. The steady state velocity is higher for [110] nanowires as explained in the text. }
\label{fig:5}
\end{figure}

\end{document}